\documentclass[10pt,twocolumn]{revtex4}
\usepackage{graphicx}
\usepackage{amsmath}
\pdfoutput=1
\newcommand{\bR}{{\bf R}}

\newcommand{\br}{{\bf r}}

\begin{document}

\title{QWalk: A Quantum Monte Carlo Program for Electronic Structure}
\author{Lucas K. Wagner\footnote{Current address: 366 Le Conte Hall \#3700; Berkeley, CA 94720}, 
Michal Bajdich, Lubos Mitas}

\affiliation{ Center for High Performance Simulation and
Department of Physics,
North Carolina State University, Raleigh, NC 27695.}

\date{\today}

\begin{abstract}
We describe QWalk, a new computational package capable of performing
Quantum Monte Carlo electronic structure calculations for molecules and
solids with many electrons.  We describe the structure of the program and its implementation of Quantum Monte Carlo methods. It is open-source, licensed under the GPL, and available at the web
site http://www.qwalk.org.
\end{abstract}

\maketitle

\section{Introduction}

Solution of the stationary Schr\"odinger equation for interacting systems of
quantum particles is one of the key challenges in quantum chemistry and 
condensed matter physics. In particular, many problems in electronic structure 
of atoms, molecules, clusters and solids require the ground and excited
 eigenstates of the electron-ion
 Born-Oppenheimer Hamiltonian 
\begin{equation}
H= -\frac{1}{2} \sum_i\nabla_i^2 -\sum_{iI}\frac{Z_I}{r_{iI}} +
\sum_{i>j} \frac{1}{r_{ij}},
\end{equation}
where upper/lower cases indicate nuclei/electrons. 
Due to the Coloumb interaction, the eigenstates are very complicated functions in the 
 in 3$N_e$-dimensional space where $N_e$ is the number of electrons. 
Over the past
six decades or so, physicists and chemists have developed many powerful approaches 
and theories that attempt to solve the electronic structure problem with varying degrees
of accuracy.  Among these are the wave function methods such as Hartree-Fock (HF) and
 post Hartree-Fock (post-HF), and also methodologies which are based on functionals
of electron density such as 
Density Functional theories (DFT).  
Because none of them are exact in practice, each of these methods occupies its place in 
the computational toolbox.  DFT represents an excellent tradeoff between accuracy and 
computational
efficiency, allowing thousands of electrons to be treated, usually getting qualitative
trends correctly for many quantities and materials such as cohesive/binding energies, many
(but not all) energy 
differences between different systems, and can even be quantitatively accurate for some quantities
(such as geometries), especially for the systems of atoms from the first two rows 
of the periodic table.  Many systems and effects are, however, not accurately described
(van der Waals systems, systems with transition metal atoms, many excitations, etc.)
and require treatment of  quantum many-body effects more accurately. One
can turn to post-Hartree-Fock methods based on sophisticated expansions of wave functions
in one-particle basis sets. These methods can be made formally exact,
unfortunately, the computational cost is substantial and the most accurate approaches 
 scale quite poorly with the system size, say, O($N_e^{5-7}$). It is very difficult to find
a method that scales well, at most O($N_e^3$), and also 
offer higher accuracy than DFT.

Quantum Monte Carlo methods fill this gap by using stochastic algorithms to treat
the many-body wave function in the full 3$N_e$-dimensional space.  It has several 
advantages--good scaling in the number of electrons (O($N_e^{2-3}$), depending on the quantity
of interest) and is amenable to 
parallel implementations at 99\% efficiency.  Over the past $\sim$20 years,
QMC has been applied to a host of systems such as model systems, atoms, molecules
 and solids, 
with impressive accuracy across this wide range~\cite{Foulkes_review,jeff_benchmark}.
For extended systems, particularly, it is the most accurate method available for total energies on the 
materials that have been tested.  Since these calculations represent rather 
recent developments, the packages for QMC are currently in development and only few 
options are available for the community at large.
We have developed a new program QWalk for general purpose QMC calculations
 written in C++ with modern programming techniques and incorporating state of the art algorithms 
in a fast and flexible code.  QWalk has already been used in several 
publications~\cite{jeff_md,michal_prl,michal_prb,pavel_1d,lucas_tmo_mol,michal_pfaff_prb}, and we would like to present a summary of its 
current capabilities.

\section{Method}

\subsection{Variational Monte Carlo}
\label{sec:qmc_vmc}
The expectation value for
an arbitrary operator $\cal O$
 and a given trial variational wave function $\Psi_T$ is given by
$$
\langle {\cal O} \rangle = \frac{\langle \Psi_T | {\cal O} | \Psi_T \rangle } {\langle \Psi_T | \Psi_T \rangle} =
\frac {\int \Psi_T^2(\bR) [{\cal O} \Psi_T(\bR) / \Psi_T(\bR)] d\bR }  {\int \Psi_T^2(\bR)  d\bR}
$$
where $\bR =(\br_1,\br_2,...,\br_{N_e})$ denotes a set of $N_e$ electron coordinates
 in 3D space. Typically, such integrals are evaluated by reducing the multi-dimensional
integral into a sum of products of low-dimensional integrals. 
Unfortunately,  this either restricts the functional form of $\Psi_T(R)$ 
or makes the calculations undo-able for more than a few electrons.
One of the key
motivations for employing stochastic approaches is to eliminate
this restriction and to gain qualitatively 
new variational freedom for describing many-body effects.

In order to evaluate the expectation value integral stochastically we first generate
a set $\{\bR_m\}$ of statistically independent sampling
points distributed according to
$\Psi_T^2(\bR)$ using the Metropolis algorithm.
The expectation value is then estimated by averaging
over the samples
$\{\bR_m\}$. For example, the VMC energy is given by the average
of the quantity called local energy
\begin{align*}
E_{VMC} &=  \frac{1}{M}\sum_{m=1}^M \frac{H \Psi_T(\bR_m)}{\Psi_T(\bR_m)}
 +\varepsilon  \\
&= \frac{1}{M} \sum_{m=1}^M E_{loc}(\bR_m) + \varepsilon
\end{align*}
with the statistical error $\varepsilon$ proportional to
$1/\sqrt{M}$. 

It is straightforward to apply the variational theorem in this framework. 
Consider a variational wave function $\Psi_T(R,P)$, where $R$ is the set 
of all the electron positions and $P$ is the set of variational parameters 
in the wavefunction  
\begin{equation}
 E(P)=\frac{ \int \Psi_T(\bR,P) H \Psi_T(\bR,P) d\bR } {\int \Psi_T^2(\bR,P) d\bR }
\end{equation}
A (hopefully) good approximation to the ground state is then the wavefunction with the set
of parameters $P$ that minimizes $E(P)$.
The stochastic method of integration
 allows us to use explicitly correlated
trial wave functions such as the Slater-Jastrow form, along with other functional forms as 
explained later.  In fact, as long as the trial function and its derivatives can be
evaluated quickly, any functional form can be used.

Within the program, this procedure is broken down into two parts: 
sampling $\Psi_T^2$ while evaluating energy and other properties, and optimizing the wave function.
The first part, sampling $\Psi_T^2$, is carried out using the Metropolis-Hastings~\cite{metropolis,hastings} algorithm.
We start with a point $\bR$ in $3N_e$ dimensional space and generate a second point $\bR'$
according to the transition probability $T(\bR'\leftarrow \bR)$.  $T$ is a completely arbitrary 
function as long as $T(\bR' \leftarrow \bR)\neq 0 \Leftrightarrow T(\bR \leftarrow \bR')\neq 0$; that is,
all moves are reversible.
We then accept the move with probability 
\begin{equation}
a=min\left(1,\frac{\Psi_T^2(\bR')T(\bR' \leftarrow \bR)}{\Psi_T^2(\bR)T(\bR \leftarrow \bR')} \right).
\end{equation}
After a few steps, the distribution converges to $\Psi_T^2$, and we continue making the moves 
until the statistical uncertainties are small enough. For atoms with effective core potentials, 
we use the moves as outlined in Ref.~\cite{Foulkes_review},
 modified with a delayed rejection step similar to Ref.~\cite{bressanini:3446}, 
although developed independently, and for full-core calculations, we use the 
accelerated Metropolis method from Ref.~\cite{unr}.  The total energy and its components are 
evaluated, as well as other properties.

We then optimize the wave function using a fixed set of sample points.  Since the 
samples are then correlated, small energy differences can be determined
with much greater precision than the total energy.
There are many quantities other than energy that, upon being minimized, will 
provide a good approximation
to the ground state wave function.  One important one is the variance of the local 
energy; that is
\begin{equation}
\sigma^2=\frac{\int d\bR \Psi_T^2(\bR) (E_{loc} - \langle E_{loc} \rangle)^2 }{ \int d\bR \Psi_T^2(\bR)}.
\end{equation}
Since $E_{loc}$ is a constant when $|\Psi_T\rangle=|\Phi_0\rangle$, the variance
will go to zero for an exact eigenstate.  There are several other possible functions, listed in 
Sec.~\ref{qwalk_optimization}, but variance and energy are the most common quantities
to minimize.

\subsection{Projector Monte Carlo}\label{subsec:PMC}

To obtain accuracy beyond a given 
variational ansatz, we employ another method which projects out the ground state
of a given symmetry from any 
trial wave function. To do this, we simulate the action of the operator 
$e^{-(H-E_0)\tau}$ on the trial function, where $\tau$ is the projection time 
and $E_0$ is the self-consistently determined energy of the ground state.  
As $\tau \rightarrow \infty$, $e^{-(H-E_0)\tau} \Psi_T \rightarrow \Phi_0$,
 where $\Phi_0$ is the ground state.  For large $\tau$, there is no general
expansion for $e^{-(H-E_0)\tau}$, but for small $\tau$, we can 
write the projection operator in $\bR$-representation as 
\begin{align*}
G(\bR',\bR,\tau) & \simeq  exp(-(\bR'-\bR)^2/2\tau) \\
 & \times exp(-\frac{\tau}{2}(V(\bR)+V(\bR')-2E_0))
\end{align*}
 which can be interpreted as a dynamic diffusion kernel $G_D(\bR',\bR,\tau)=exp(-(\bR'-\bR)^2/2\tau)$ 
times a branching kernel
$G_B(\bR',\bR,\tau)=exp(-\frac{1}{2}(V(\bR)+V(\bR')-2E_0))$.

The basic idea of projector Monte Carlo is to sample a path 
$G(\bR_N,\bR_{N-1},\tau)...G(\bR_2,\bR_1,\tau)\Psi_T(\bR_1)$.  For $N$ large enough (for a long 
enough path), the distribution of $\bR_N$ will approach $\Phi_0$. 
However, to interpret this as a stochastic process, the path distribution must be positive; that is, the product of all $G$'s with 
$\Psi_T$ must be positive.  This gives rise to the fixed node approximation~\cite{jbanderson75,jbanderson76,moskowitz82,reynolds82}, 
where the nodes (the places where the trial function equals zero) 
of the trial wave function are used as approximation to 
the nodes of the ground state wave function. One can avoid this restriction by performing 
a released-node calculation~\cite{ceperley_adler}, although the price is a change from polynomial 
to exponential scaling with system size.  With the nodal 
constraint, the projector Monte Carlo approach typically obtains 90-95\% of the correlation 
energy in an amount of time proportional to  $N_e^{\alpha}$ where $\alpha=2,3$ depending 
on actual implementation and type of the system. 
In what follows we therefore assume that the fixed-node condition is enforced
and therefore $\Phi_0$ is the antisymmetric ground state for
a given fixed-node boundary condition.

In actual calculations, we perform an importance-sampling transformation, 
where $G(\bR',\bR,\tau)$ is replaced 
by the importance sampled Green's function 
\begin{equation}
\tilde G(\bR',\bR,\tau)=\Psi_T(\bR') G(\bR',\bR,\tau)/\Psi_T(\bR)
\end{equation} The dynamic part of the Green's function then becomes
\begin{equation}
G_D(\bR',\bR,\tau)=exp(-(\bR'-\bR-\tau\nabla ln\Psi_T(\bR))^2/2\tau)
\end{equation}
and the branching part becomes
\begin{equation}
G_B(\bR',\bR,\tau)=exp(-\frac{1}{2}(E_L(\bR)+E_L(\bR')-2E_0)),
\end{equation}
both of which are much better-behaved stochastically, since the 'force' $\nabla ln\Psi_T(\bR)$ 
biases the walk to where
the wavefunction is large, and the local energy $E_L(\bR)$ is much smoother than the 
potential energy.
Then if we 
generate the path $\tilde G(\bR_N,\bR_{N-1},\tau)...\tilde G(\bR_2,\bR_1,\tau)\Psi_T^2(\bR_1)$,
 for large enough $N$, 
the distribution of $\bR_N$ is $\Psi_T(\bR_N)\Phi_0(\bR_N)$, which is called the mixed
 distribution.  
The ground state energy is obtainable by evaluating the integral
 $\int \Psi_T \Phi_0 H\Psi_T/\Psi_T d\bR = \int \Phi_0 H \Psi_T d\bR= E_0$,
 since $\Phi_0$ is an eigenstate of $H$ within the nodal boundaries.  In QWalk, two versions of the projector method are implemented: 
Diffusion Monte Carlo, which has the advantage that the large $N$ limit is easily obtained, and Reptation Monte Carlo, which makes the 'pure' distribution $\Phi_0^2$ available.

Diffusion Monte Carlo has been discussed by many authors~\cite{Foulkes_review,unr}, and suffice it to say that it attains the mixed distribution by starting with a distribution of $\Psi_T^2$ and interpreting the action of the Green's function as a stochastic process with killing and branching, eventually ending up with $\Psi_T \Phi_0$.  It has the advantage that the $\tau \rightarrow \infty$ limit is easy to achieve, but the disadvantage of not having access to the pure distribution.  A more subtle limitation is that the branching process spoils any imaginary-time data and can decrease the efficiency of the simulation if there is too much branching.  Even with these limitations, in current 
implementations DMC is probably the most efficient way to obtain the fixed-node approximation to the ground state energy.

For quantities that do not commute with the Hamiltonian, we use Reptation Monte Carlo~\cite{Baroni_RMC} with the bounce algorithm~\cite{pierleoni_rmc}. 
We sample the path distribution
\begin{equation}
\Pi(s)=\Psi_T(\bR_0) G(\bR_0,R_1,\tau)\ldots G(\bR_{n-1},\bR_n,\tau) \Psi_T(\bR_n)
\end{equation}
where $s=[\bR_0, \bR_1, \ldots,\bR_{n-1}, \bR_n]$ is a projection path.
In the limit as $\tau \rightarrow \infty$, $exp(-H\tau) | \Psi_T \rangle \rightarrow | \Phi_0 \rangle$,
the ground state,
and, since it is a Hermitian operator, the conjugate equation also holds. Therefore, 
the distribution of $\bR_0$ and $\bR_n$ is the mixed distribution $\Psi_T(\bR) \Phi_0(\bR)$,and 
the distribution of $\bR_{n/2}$ is
$\Phi_0^2(\bR_{n/2})$ in the limit as $n \rightarrow \infty$.
We evaluate the energy as $E_{RMC}=\langle [E_L(\bR_0)+E_L(\bR_N)]/2 \rangle$
and operators non-commuting with $H$ as $O_{RMC}=\langle O(\bR_{N/2}) \rangle $
Reptation Monte Carlo does not include branching, 
instead it uses an acceptance/rejection step. 
 This is a tradeoff, allowing us to project only for 
a finite $\tau$, since otherwise the probability distribution function 
is not normalizable, but allowing access to the pure distribution 
and imaginary time correlations.  The path can sometimes get stuck due to rejections 
 even with the bounce algorithm, which is a well-known  limit on  the efficiency 
of the algorithm.  In QWalk, RMC is approximately as efficient as DMC
until the rejection rate begins to increase, making the path move very 
slowly.  In our current implementation we empirically find that this slowdown
occurs at approximately 150 electrons, although it also depends on the quality
of the trial wave function.

\section{Organization and Implementation}

\begin{figure*}
\includegraphics[width=\textwidth]{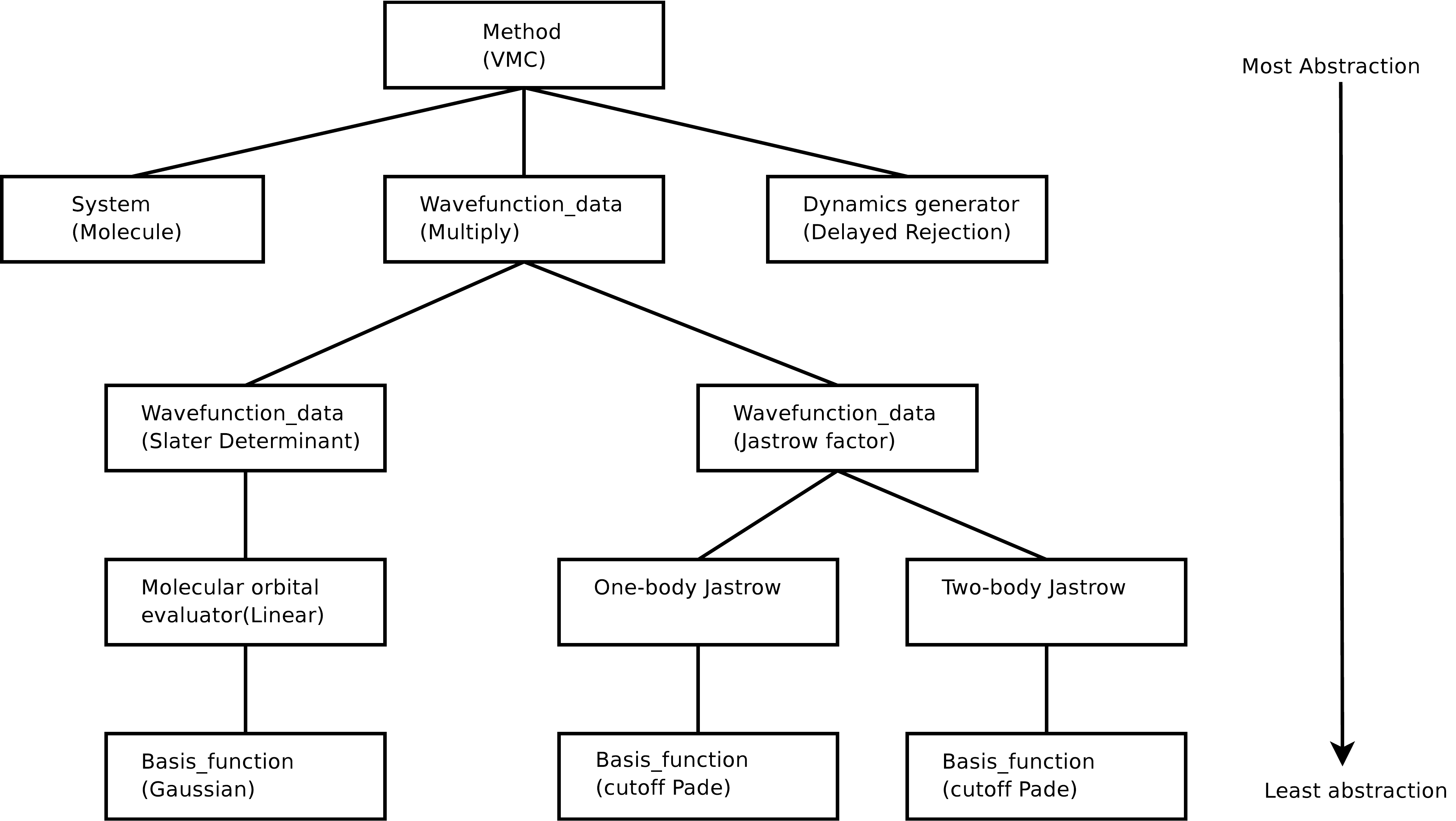}
\caption{Calculation structure for the VMC method on a molecule using a 
Slater-Jastrow wave function. }
\label{fig:tree}
\end{figure*}

The code is written in a combination of object-oriented and procedural techniques. 
The object-oriented approach is coarse-grained, creating independent sections of code
that are written efficiently in a procedural fashion.
 It is extremely modular; almost every piece can be removed and replaced with another.
 A contributor of a module only has to change one line in the main code to allow use of a new module. 
This allows for flexibility while keeping the code base relatively simple and separable.  
The modular structure also allows for partial rewrites of the code without worrying 
about other parts.  In fact, each major module has been rewritten several times in this 
manner as we add new features and refactor the code.  For the user, this structure 
shows itself in flexibility.

\begin{table}
\caption{The central objects of the code and their physical correspondents}
\label{table:correspondence}
\begin{center}
\begin{tabular}{ll}
\hline
{\bf Module name} & {\bf Mathematical object} \\
\hline
System & parameters and form of the Hamiltonian \\
Sample point & $\bR$, the integration variables \\
Wave function type  & Wave function ansatz \\
Wave function & $\Psi_T(\bR)$, $\nabla\Psi_T(\bR)$, $\nabla^2\Psi_T(\bR)$ \\
Dynamics generator & Metropolis trial move \\
                   & (Green's function) \\
\hline
\end{tabular}
\end{center}

\end{table}

The modules form a tree of successive abstractions (Fig.~\ref{fig:tree}).  At the top of the tree
is the QMC method, VMC in this case.  It works only in terms of the objects directly
below it, which are the concepts of System, Wave function data, etc. (see Table~\ref{table:correspondence}).
These in turn may have further abstractions below them, as we've shown for the wave function
object.  The highest wave function object is of type `Multiply', which uses two 
wave function types to create a combined wave function.  In this case, it multiplies 
a Slater determinant with a Jastrow correlation factor to form a Slater-Jastrow
function.  Since the wave functions are pluggable, the Slater determinant can be
replaced with any antisymmetric function, as well as the Jastrow factor.  
The type is listed along with the specific instant of that type in parenthesis.  At 
each level, the part in parenthesis could be replaced with another module
of the same type.

\begin{figure}
\hrule
\footnotesize
\begin{verbatim}
Vmc_method::run(vector <string> & vmc_section,
                vector <string> & system_section,
                vector <string> & wavefunction_section) {

  //Allocate the objects we will be working with
  System * sys=NULL;
  allocate(sys, system_section);

  Wavefunction_data * wfdata=NULL;
  allocate(wfdata, sys, wavefunction_section);

  Sample_point * sample=NULL;
  sys->generateSample(sample);
  Wavefunction * wf=NULL;
  wfdata->generateWavefunction(wf);

  //the Sample_point will tell the Wavefunction
  //when we move an electron
  sample->attachWavefunction(wf);
  sample->randomGuess();

  //This is the entire VMC algorithm
  for(int s=0; s< nsteps; s++) { 
    for(int e=0; e < nelectrons; e++) { 
       dynamics_generator->sample(e,timestep,wf,sample);
    } //end electron loop
    //gather averages
  } //end step loop

  //report final averages
\end{verbatim}
\hrule
\caption{Simple VMC code}
\label{fig:vmc_code}
\end{figure}
We present an implementation of the VMC algorithm as an example of how the code is 
organized (Fig.~\ref{fig:vmc_code}).  For reasons of space, we do not write the function line-by-line, which includes 
monitoring variables, etc., but instead give a sketch of the algorithm.  The VMC method works at the highest level of 
abstraction, only in terms of the wave function, system, and random dynamics.  It does not care what kind of
system, wave function, etc. are plugged in, only that they conform to the correct interfaces. In Appendix~\ref{app:qwalk_module}, 
we give an example of how to create a new module.

We will now provide a listing of the available modules for the major types, along with some details of their
implementation.

\section{Methods}

\subsection{Variational Monte Carlo }

The VMC module implements the Metropolis method to sample the probability density
$\Psi_T^2(\bR)$.  It has been described in Sec.~\ref{sec:qmc_vmc} to some detail--the method
is more or less a direct translation.  Beyond the basic algorithm, it implements correlated sampling as explained 
in Sec.~\ref{correlated_sampling_basic}
for small energy differences between very similar systems.

\subsection{Optimization of Wave Functions}

\label{qwalk_optimization}

We have implemented three different methods for optimization.  
All methods are capable of optimizing the first three objective functions from Table~\ref{table:optimization}.  
In principle, any objective function from this table will obtain the correct ground state 
with an infinitely flexible function, but may obtain 
different minima for incomplete wave functions and some are easier to optimize than others.  
The first (OPTIMIZE) is based on Umrigar 
et al.'s~\cite{umrigar_varopt} variance optimization.  The method minimizes the objective function
on a set of fixed configurations from VMC using a conjugate gradient technique, usually not 
reweighting the averages as the wave function changes.  Optimizing the energy using 
OPTIMIZE is quite expensive, because it requires many configurations to evaluate an 
unbiased estimate of the energy derivative.

The next two are based on Umrigar and Filippi's Newton 
optimization~\cite{umrigar_optimization2} method.  OPTIMIZE2 also uses a fixed set of configurations,
but instead of evaluating only the first derivatives of the objective function, as conjugate gradients
do, it uses a low-variance estimator for the Hessian matrix and Newton's method to find the zeros 
of the first derivatives.  OPTIMIZE2 is able to produce better wave functions with lower energies
than OPTIMIZE by directly optimizing the energy even for very large systems (we have applied it for
up to 320 electrons) while costing slightly more.  

Finally, NEWTON\_OPT uses a fixed set of configurations to calculate the 
same low-variance estimator for the Hessian matrix only at the single step, 
then evaluates the optimal length of the optimization step 
using VMC correlated sampling~\cite{umrigar_optimization2}. 
The later step enables us to decrease the number of iterations needed to converge.  
Further, this method is able to find the very lowest energy wave function, 
since the configurations are regenerated at every optimization step.
However, the expense of one iteration in NEWTON\_OPT is larger than 
for other two methods due to the additional cost associated with VMC and VMC correlated sampling.

\begin{table}
\caption{Optimization objective functions implemented }
\label{table:optimization}
\begin{center}
\begin{tabular}{lr}
\hline
{\bf Function} & {\bf Minimized quantity} \\
\hline
Variance & $\langle(E_L(\bR)-E_{ref})^2 \rangle$ \\
Energy & $\langle E_L(\bR) \rangle$\\
Mixed  & $aEnergy+(1-a)Variance$,$0<a<1$  \\
Absolute value & $\langle |E_L(\bR)-E_{ref}| \rangle$ \\
Lorentz & $\langle ln(1+(E_L(\bR)-E_{ref})^2/2) \rangle$ \\
\hline
\end{tabular}
\end{center}

\end{table}

\subsection{Diffusion Monte Carlo}

DMC is implemented almost identically to VMC, except that the time step is typically
much smaller and each walker accumulates a weight equal to 
$\exp(-\frac{\tau_{eff}}{2}(E_L(\bR')+E_L(\bR)-2E_{ref}))$.
Since we use an acceptance/rejection step, $\tau_{eff}$ is chosen somewhat 
smaller than $\tau$ as $\tau_{eff}=p\tau$, where $p$ is the acceptance ratio.
 To control the fluctuations in the weights, we employ a constant-walker
branching algorithm, which improves the parallel load balancing properties of 
DMC.  Every few steps we choose a set of walkers that have 
large weights ($w_1$) for branching.  Each one of these walkers is matched with a 
smaller weight walker ($w_2$) which is due for killing.  The large weight walker is
branched and the small weight walker is killed with probability $\frac{w_1}{w_1+w_2}$, with each copy gaining 
a weight of $\frac{w_1+w_2}{2}$.  Otherwise, the small weight walker is branched 
and the large weight walker is killed, with the copies having the same weight as
before.  Walkers are then exchanged between nodes to keep the number of walkers on each node 
constant, and thus preserve high parallel efficiency. QWalk keeps track of two numbers: $E_{ref}$ and $E_0$.  $E_{ref}$ is first set to the VMC 
average energy, and then to the energy of the last block.  The energy that goes into the 
weights, $E_0$, is then calculated every few steps as 
\begin{equation}
E_0=E_{ref}-log\left(\frac{\sum w_i}{N_{conf}}\right),
\end{equation}
where $N_{conf}$ is the number of sample points (configurations) in the simulation.

During the DMC calculation, the local energy will very occasionally fluctuate 
down significantly, causing the 
weight to increase too much. Of course, this is very much
dependent on the quality of the trial function and the studied system.
  This can be fixed by cutting off the weights.
For fluctuations beyond ten standard deviations of the energy, we smoothly 
bring the effective time step to zero for the weights, which avoids the 
efficiency problem without introducing a noticeable error. 
The bias due to this cutoff goes to zero as the time step goes to zero or
as the trial function approaches the exact one.

\subsection{Reptation Monte Carlo}

The fluctuations in the local energy part of the Green's function
can cause the path in RMC to get stuck, so we cut off the effective
time step in the same way as in DMC.  The branching part of the Green's function
is otherwise quite smooth.  We use the same dynamic Green's function
as we do in DMC (either a standard Metropolis rejection step or the UNR~\cite{unr}
algorithm), so we accept/reject based only on the branching part 
of the Green's function.  We use the bounce algorithm~\cite{pierleoni_rmc},
which improves the efficiency by allowing the path to explore the many-body
phase space much more quickly.

\subsection{Correlated Sampling}

\label{correlated_sampling_basic}

Correlated sampling is a technique where one samples two very similar systems with the same 
sets of samples.  The variance in the difference will decrease as $Var(X-Y)=Var(X)+Var(Y)-2Cov(X,Y)$,
so for perfectly correlated sampling, the variance will be zero for the difference.  In
QWalk, this is handled by performing a primary walk that samples some probability 
distribution $P_1(X)$.  Averages are obtained as usual by calculating the integral
$\langle O_1 \rangle =\int P_1(X) O_1 dX$.  Suppose we wish to find $\langle O_2-O_1 \rangle$.
It can be written as 
\begin{equation}
\int P_2(X)O_2 - P_1(X)O_1 dX= \int P_1(X)\left[\frac{P_2}{P_1}O_2 -O_1\right] dX.
\end{equation}
Since we are sampling $P_1(X)$, in the Monte Carlo averaging, this integral is evaluated by averaging the 
weighted difference over sample points:
\begin{equation}
\sum_i^N \left[ \frac{w_i(X_i) O_2(X_i)}{\sum_j w_i(X_i)} -\frac{O_1(X_i)}{N}\right]
\end{equation}
The difference in the methods is only in how they determine the weights.

VMC, DMC and RMC all support correlated sampling between arbitrary systems.  In VMC,
the weights are $w(X)=\frac{\Psi_2^2(X)}{\Psi_1^2(X)}$, which is an exact relationship.
DMC and RMC both require some approximation to the Green's function to weight 
the secondary averages properly.  In both, we use the approximation of Filippi and Umrigar~\cite{filippi_force},
who discuss the subject in a greater detail.

\section{Systems}

\subsection{Boundary Conditions}

Most systems of interest are treatable either by open boundary conditions or 
periodic boundary conditions.  Adding new boundary conditions is also
quite simple.
Molecules with arbitrary atoms, charge, spin state, and with finite electric field are supported.
In 3D periodic systems, the calculation can be done at any real k-point, allowing k-point integrations.
In many-body simulations, there is an additional finite size approximation due to the Coulomb
interaction between image electrons.  We correct this as $\delta E = \frac{c}{r_s}$, where 
the $r_s$ is that of the homogeneous electron gas and $c$ has been empirically fitted 
to 0.36 Hartrees.  We have found this correction to function about as well as other 
attempts to correct the finite size error~\cite{cambridge_mpc,chiesa_finite_size}.
 The code has been used on systems with up to 135 atoms and 1080 electrons; the limiting factor
is the amount of computer time needed to reduce the stochastic uncertainties.

\subsection{Pseudopotentials}

QWalk accepts pseudopotentials as an expansion of nonlocal angular momentum operators:
\begin{equation}
\hat{V}_{ECP}=V_{local} (\bR) + \sum_{l=0}^{lmax} V_l(\bR) |l\rangle\langle l|
\end{equation}
for arbitrary maximum angular moment.  $V_l$ is a basis function object that is typically a 
spline interpolation of a grid or a sum of Gaussian functions.
While any pseudopotential of this form can be used, we use soft potentials in which the $\frac{Z}{r}$ divergence has been removed from the nuclei-electron interaction.  These potentials have been created 
specifically for QMC and are available in the literature~\cite{Lester_psp,Lee_psp,Trail_psp,Dolg_psp_qmc}, although more traditional
Hartree-Fock or DFT pseudopotentials in the Troullier-Martins form work as well.

\section{Forms of the Wave function}

For chemical problems, the first-order trial function is usually written as a single Slater determinant of Hartree-Fock or Density 
Functional Theory orbitals multiplied by a correlation factor (known as a Jastrow factor) which is optimized in Variational Monte Carlo. 
Between 90\% and 95\% of the correlation energy is typically obtained with this trial wave function in Diffusion Monte Carlo.  

One of the attractions of QMC is that, since all the integrals are done by Monte Carlo, almost any ansatz can be used, as long as it is reasonably quick to evaluate.
QWalk's  modular structure makes adding new wave function forms as simple as coding one-electron updates of the function value and derivatives, 
and adding one line to the main code to support loading of the module.  We have implemented several forms of wave functions, which the user 
can combine.  For example, to create the Slater-Jastrow wave function, the user first asks for a multiply object, which contains two wave 
function objects.  The user then fills in a Slater determinant object and a Jastrow object. For a Pfaffian-Jastrow wave function, the user 
replaces the Slater determinant input with the Pfaffian input.  Obviously, it is up to the user to make sure that the total wave function 
is properly antisymmetric and represents the problem correctly. 

\subsection{Slater Determinant(s)}

This is the standard sum of Slater determinants, 
written as $\Psi_T=\sum{c_i D_i^\uparrow D_i^\downarrow}$, where $D_i^{\uparrow(\downarrow)}$ 
is a determinant of the spin up (down) one-particle orbitals.  
The weights of the determinants $\{c_i\}$ are optionally optimizable within VMC. 

\subsection{Jastrow Factor}

The Jastrow factor is written as $e^U$, where
\begin{multline}
\label{eqn:jastrow_factor}
U=\sum_{iIk} c_k^{ei} a_k(r_{iI}) + \sum_{ijk} c_k^{ee} b_k(r_{ij}) \\
+ \sum_{ijIklm} c_{klm}^{eei} [a_k(r_{iI})a_l(r_{jI})+a_k(r_{jI})a_l(r_{iI})]b_m(r_{ij}),
\end{multline}
$i,j$ are electron indexes, and $I$ is a nuclear index. Both the coefficients and parameters within the basis functions can be optimized.  
In addition, the $\{c^{ee}\}$ and $\{c^{eei}\}$ coefficients can be made spin-dependent. 
For the basis functions, we satisfy the exact electron-electron cusp conditions with the function $b(r)=cp(r/rcut)/(1+\gamma p(r/rcut)) $, where
$ p(z)=z-z^2+z^3/3 $, $ \gamma $ is the
curvature, which is optimized, and $c$ is the cusp(1/4 for like spins and 1/2 for unlike spins).
Further correlation is added by including functions of the form 
$ b_k(r)=a_k(r) = \frac{1-zpp(r/rcut)}{1+\beta zpp(r/rcut)}$ where $zpp(x)=x^2(6-8x+3x^2)$ and $\beta$ is an optimized parameter.  These functions have several favorable properties, going smoothly to zero at a finite cutoff radius, and covering the entire functional space between 0 and $rcut$. This allows the Jastrow factor to be very compact, 
typically requiring optimization of around 25 parameters while still coming close to saturating the functional form.  While these are the standard basis functions, they can be replaced or augmented by any in the program by
a simple change to the Jastrow input.  The third term in Eq.~(\ref{eqn:jastrow_factor}), which sums over two electron 
indexes and ionic indexes, can be expensive to evaluate for large systems and is sometimes
excluded.  A Jastrow factor with only the first two terms is called a two-body Jastrow, and
with the $eei$ term included is called a three-body Jastrow.

\subsection{Pfaffian Pairing Wave Function}

Pairing wave functions with a Jastrow factor for molecules were first investigated by Casula and coworkers~\cite{casula-2004-121}, 
who studied the constant number of particles projection of the BCS wave function.  
The general Jastrow-BCS pairing wave function can be expressed as $\Psi_T=e^U{\rm det}[{\boldsymbol \Phi}] $, 
where $e^U$ is the  Jastrow factor of above and the matrix ${\boldsymbol \Phi}_{ij}=\phi(r_i,r_j)$ is the pairing function between 
opposite-spin electrons (the function is easily extended for $N_{up} \neq N_{down}$).  This function
contains the Slater determinant as a special case (for singlet spin state) when $\phi$ 
is written as the sum over the occupied single-particle orbitals: $\phi(r_i,r_j)=\sum_k^{N_e} \varphi_k(r_i)\varphi_k(r_j)$.  
We have implemented the Pfaffian~\cite{michal_prl} pairing wave function, which allows not only unlike-spin pairing, 
as the canonical projection of the BCS wave function does, but also allows like-spin pairing.  
The general Pfaffian pairing wave function $\Psi_{PF}$ is written as the Pfaffian of the antisymmetric matrix
\begin{equation}
\Phi_{PF}={\rm pf}\left[ \begin{array}{ccc}
{\boldsymbol \xi}^{\uparrow\uparrow} &
{\boldsymbol \Phi}^{\uparrow\downarrow} &
 {\boldsymbol\varphi}^{\uparrow} \\
-{\boldsymbol \Phi}^{\uparrow\downarrow T} &
{\boldsymbol \xi}^{\downarrow\downarrow} &
{\boldsymbol \varphi}^{\downarrow} \\
-{\boldsymbol\varphi}^{\uparrow T} &
 -{\boldsymbol\varphi}^{\downarrow T} &
 0 \;\;
\end{array}
\right],
\end{equation} 
where ${\boldsymbol \Phi}^{\uparrow\downarrow}$,  ${\boldsymbol \xi}^{\uparrow\uparrow(\downarrow\downarrow)}$  
and ${\boldsymbol \varphi}^{\uparrow(\downarrow)}$ represent the following block matrices. 
The ${\boldsymbol \Phi}^{\uparrow\downarrow}$ is the singlet pairing matrix from above BCS wave function, 
the ${\boldsymbol \varphi}^{\uparrow(\downarrow)}$ includes additional unpaired one-particle orbitals for a spin-polarized system.  
Finally, the ${\boldsymbol \xi}^{\uparrow\uparrow(\downarrow\downarrow)}$ are antisymmetric triplet pairing matrices.
The operation of the Pfaffian ensures that the entire wave function is antisymmetric.
The Pfaffian wave function contains the BCS wave function as a special case without triplet pairing, and
therefore contains the Slater determinant wave function as well. The general expansion for $\Phi$ is
\begin{equation}
\Phi^{\uparrow\downarrow(\br_1,\br_2)}=\sum_{kl} c_{kl} \varphi^{\uparrow}_k(\br_1)\varphi^{\downarrow}_l(\br_2)
\end{equation}
under the constraint that $c_{kl}=c_{lk}$.  ${\boldsymbol \xi}$ is written in a very similar way:
\begin{equation}
{\boldsymbol \xi}^{\uparrow\uparrow(\downarrow\downarrow)}(\br_1,\br_2)=\sum_{kl} d_{kl}^{\uparrow\uparrow(\downarrow\downarrow)} 
\varphi^{\uparrow(\downarrow)}_k(\br_1)
\varphi^{\uparrow(\downarrow)}_l(\br_2)
\end{equation}
under the constraint that $d_{kl}^{\uparrow\uparrow(\downarrow\downarrow)}=-d_{lk}^{\uparrow\uparrow(\downarrow\downarrow)}$.  
The sum extends over the space of occupied and virtual orbitals. 
All pairing functions as well as unpaired orbitals are fully optimizable within VMC method.
The extensions of Pfaffian pairing wave function to linear combinations of Pfaffians 
with one or many sets of different pairing functions are also fully implemented.
For more information about the performance and implementation of the Pfaffian wave function, 
see Refs.~\cite{michal_prl,mb_thesis}.

\subsection{Backflow Correlated Wave Function}

Another way to systematically improve the nodal structure of trial 
wave function is through the introduction of backflow transformation~\cite{feynman,schmidt_bf,panoff,moskowitz,kwon1,kwon2,kwon3,drummond_bf,rios_bf}. 
Given a trial wave function of form $\Psi_T({\bf R})=\Psi_A({\bf R})\times \exp[U({\bf R})]$,
the nodal structure is completely defined by the nodes of its antisymmetric part $\Psi_A({\bf R})$.
The backflow transformation replaces $\Psi_A({\bf R})$ by $\Psi_A({\bf X})$, where ${\bf X}=({\bf x}_1,{\bf x}_2,\ldots)$ are some quasi-coordinates 
dependent on all electron positions ${\bf R}$, such the overall antisymmetry is preserved. 
The nodes of $\Psi_A({\bf X})$ can then differ from nodes of $\Psi_A({\bf R})$ and 
improve the fixed-node approximation.

The QWalk implementation of the backflow transformation into Slater and Pfaffian wave functions closely follows the approach in 
Refs.~\cite{kwon2,rios_bf}. The quasi-coordinate of $i$th electron at position ${\bf r}_i$ is given as 
\begin{align}
{\bf x}_i&={\bf r}_i+{\boldsymbol \xi}_i({\bf R}) \nonumber \\
&={\bf r}_i+{\boldsymbol \xi}_i^{en}({\bf R})+{\boldsymbol \xi}_i^{ee}({\bf R})+{\boldsymbol \xi}_i^{een}({\bf R}),
\end{align}
where ${\boldsymbol \xi}_i$ is the $i$th electron's backflow displacement 
divided to the contributions from one-body (electron-nucleus), two-body (electron-electron) 
and three-body (electron-electron-nucleus) terms. 
They can be further expressed as 
\begin{align}\label{eg:bfterms}
{\boldsymbol \xi}_i^{en} ({\bf R})&=\sum_I \left [\sum_{k} c_k^{ei} a_k(r_{iI})\right] {\bf r}_{iI}\\
{\boldsymbol \xi}_i^{ee} ({\bf R})&= \sum_{j\ne i} \left [\sum_{k} c_k^{ee} b_k(r_{ij}) \right] {\bf r}_{ij} \\
{\boldsymbol \xi}_i^{een}({\bf R})&= \nonumber \\
\sum_{I,j\ne i} &\left[ \sum_{klm} c_{klm}^{eei} \left[a_{k,r_{iI}}a_{l,r_{jI}}+a_{k,r_{jI}}a_{l,r_{iI}}\right] b_{m,r_{ij}} \right]  {\bf r}_{ij} \nonumber \\
               +&\left[ \sum_{klm} d_{klm}^{eei} \left[a_{k,r_{iI}}a_{l,r_{jI}}+a_{k,r_{jI}}a_{l,r_{iI}}\right] b_{m,r_{ij}} \right]  {\bf r}_{iI},
\end{align}
where ${\bf r}_{ij}={\bf r}_i-{\bf r}_j$ and ${\bf r}_{iI}={\bf r}_i-{\bf r}_I$. 
The terms in the large square brackets are identical to our familiar one, two and two three-body Jastrow terms from Eq.~(\ref{eqn:jastrow_factor}). 
The implementation of backflow transformation therefore takes great advantage 
of already existent Jastrow. The improvement in nodal structure and gains in correlation energies 
can be achieved by optimizing all the Jastrow parameters within backflow transformation.
For more details about implementation and performance of backflow transformation in QWalk see Ref.~\cite{mb_thesis}.

\section{One-particle orbital evaluation}

We provide two major ways of evaluating the one-particle orbitals, the most expensive 
part of the QMC calculation.  For a single electron, this is the problem of finding $\vec{m}=M_{orb}\vec{b}$, where $\vec{m}$ 
is a vector of the values of each orbital, $M_{orb}$ is the orbital coefficient matrix, and $\vec{b}$ is the vector of basis functions. The first (CUTOFF\_MO) is a linear scaling technique, which, for localized
orbitals and large enough systems, will take $O(N)$ time to evaluate all orbitals
for all electrons.  For each basis function, it creates a list of orbitals for which the coefficient
is above a cutoff.  This is done at the beginning of the calculation.  Then, given an electron position,
it loops over only the basis functions within range of the electron, and then only the 
orbitals contributed to by the basis function.  These are both $O(1)$ cost for large enough
systems, so all the orbitals for each electron is evaluated in $O(1)$ time, giving $O(N)$ scaling.

The second method (BLAS\_MO) is slightly simpler.  While it scales in principle as $O(N^2)$, it can
be faster than CUTOFF\_MO in medium-sized systems and certain types of computers
that have very fast BLAS routines, such as Itaniums.  Given an electron position, it loops through the basis functions
within range of the electron, and adds to each molecular orbital the coefficient times the value 
of that basis function using fast BLAS routines.

\section{Example calculation}

\begin{figure}
\begin{center}
\includegraphics[width=\columnwidth]{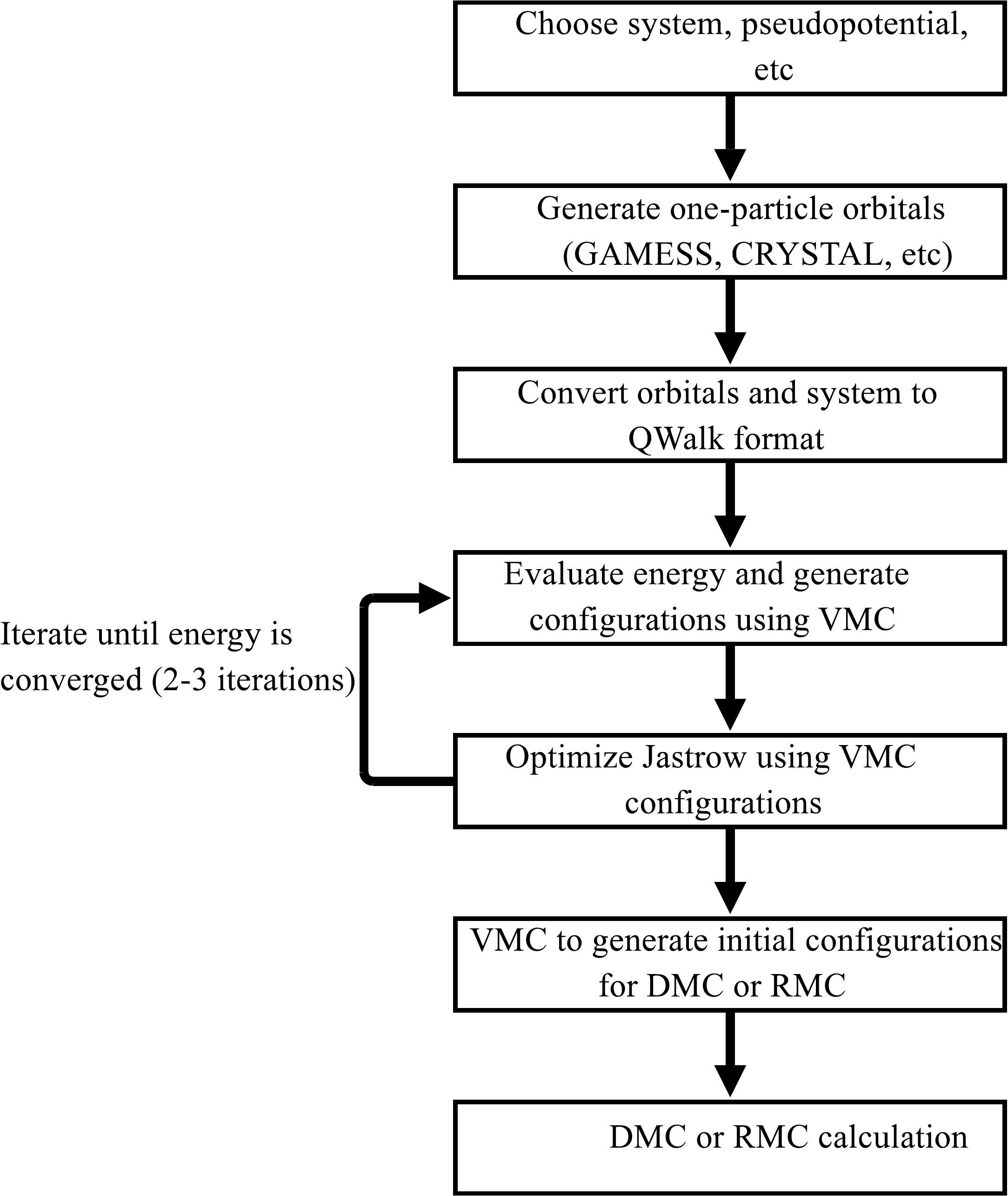}
\end{center}
\caption{Flow of a QMC calculation}
\label{fig:flowchart}
\end{figure}

To give a feeling for the flow of the program, we will go through a simple calculation.  A schematic of the procedure is 
given in Fig.~\ref{fig:flowchart}.  The first two steps are to choose the system and use a one-particle code such as 
GAMESS or CRYSTAL to prepare the one-particle orbitals, which is done as usual for the code.  The converter program included with QWalk then creates
the system, slater, and jastrow files automatically, so all the user must do is use the include directive to 
use them. In Fig.~\ref{fig:vmc}, we evaluate the properties of the starting Slater wave function by creating 500 
electron configurations, and then propagating them for 16 blocks, each of which consists of 10 Monte-Carlo steps with a time step of 1.0 a.u.  The final set of configurations is then stored in 'configfile', and QWalk outputs the total energy and other properties that have been accumulated along the way.
\begin{figure}
\hrule
\begin{verbatim}

#load the converted pseudo-nuclei, number of electrons
include sysfile   

#load the Slater determinant of one-particle orbitals
trialfunc { include slaterfile } 

#
method {  VMC
   nconfig 500            #number of configurations 
   nblock 16              #averaging blocks
   nstep 10               #steps to take per block
   timestep 1.0           #timestep to use
   storeconfig configfile #save configurations to
                          #a file
}
\end{verbatim}
\hrule
\caption{Example input file for VMC evaluation of properties.  This corresponds to the fourth box in Fig.~\ref{fig:flowchart}.}
\label{fig:vmc}
\end{figure}

We then wish to obtain some correlation energy by adding the Jastrow factor (Fig.~\ref{fig:opt}).  The converter has already created a 
null Jastrow wave function, so we request a Slater-Jastrow wave function.  The first wave function is the Slater 
determinant that we used before, and the second is the Jastrow created by the converter.  We request optimization using a fixed set of walkers that we generated in the previous VMC run.

\begin{figure}
\hrule
\begin{verbatim}

include sysfile

trialfunc { 
 #a meta wave function that
 #multiplies two wave functions
  multiply 
  wf1 { include slaterfile }  

   #Jastrow correlation factor (created by converter)
  wf2 { include jastrowfile }
}

method {  OPTIMIZE
   nconfig 500          
   iterations 30        

   #read in the configurations from 
   #the previous VMC run
   readconfig configfile
}
\end{verbatim}
\hrule
\caption{Example input file for optimization of variational parameters.  This is the fifth box in Fig.~\ref{fig:flowchart}.}
\label{fig:opt}
\end{figure}

Finally, we wish to evaluate properties of the new correlated wave function using the VMC routine (Fig.~\ref{fig:vmc_opt}).  This is
the same as the last, except that we include the output wave function from the optimization in the trialfunc
section.  Also in the example, we perform a DMC calculation immediately after the VMC calculation.  Its input is nearly identical to that already discussed.
\begin{figure}
\hrule
\begin{verbatim}
include sysfile

#load the wavefunction file generated by OPTIMIZE
trialfunc { include optfile.wfout }

#Same as above 
method {  VMC
   nconfig 500 
   timestep 1.0
   nstep 10 
   nblock 16
   readconfig configfile
   storeconfig configfile
}

#perform DMC

method { DMC
   nconfig 500  
  timestep 0.02     #smaller timestep
  nstep 50          #more steps per block
  nblock 16
  readconfig configfile
  storeconfig configfile
}

\end{verbatim}
\hrule
\caption{Example input file for evaluation of properties of the correlated wave function, plus a DMC calculation.  
This corresponds to the sixth and seventh block in Fig.~\ref{fig:flowchart}.}
\label{fig:vmc_opt}
\end{figure}

\section{Other Utilities}

\subsection{Conversion of One-particle Orbitals}

Currently, QWalk can import and use the orbitals from GAMESS~\cite{gamess} (gaussian basis on molecules),
 CRYSTAL~\cite{crystal} (gaussian basis for extended systems), SIESTA~\cite{siesta}, and GP~\cite{GP} (plane waves for extended systems).
The GP interface is not currently available for distribution due to licensing issues.  More interfaces are planned, 
and are quite easy to add.

\subsection{Plane Wave to LCAO converter} 
Gaussian basis sets have been used in quantum chemistry for years and have been developed to the point that there are
well-defined sets which saturate the one-body Hilbert space surprisingly quickly.  They are localized, which improves the scaling of QMC, and allow a very compact expression of the one-particle orbitals, so less basis functions need to be calculated.  Overall, a gaussian representation can improve the performance of the QMC code by orders of magnitude over the plane-wave representation.  We have developed a simple method to do this conversion that is fast and accurate.  We start with the plane-wave representation of the $k$-th orbital $\Phi_k(\vec{r})=\sum_{\vec{G}}c_{k\vec{G}}e_{\vec{G}}(\vec{r})$,  and wish to find the LCAO equivalent
$\Phi_k^{LCAO}(\vec{r})=\sum_j a_{kj} \phi_j(\vec{r})$, where $e_{\vec{G}}$ is a plane-wave function
and $\phi_j$ is a Gaussian function.  Maximizing the overlap between $\Phi_k$ and $\Phi_k^{LCAO}$, we obtain $Sa_k=Pc_k$, where $S_{ij}=\langle \phi_i | \phi_j \rangle$ and $P_{i\vec{G}}=\langle \phi_i | e_{\vec{G}} \rangle$.
Then the Gaussian coefficients are given as $a_k=S^{-1}Pc_k$.
All the overlap integrals are easily written in terms of two-center integrals for $S$, and $P$ is easily evaluated 
in terms of a shifted Gaussian integral.  The limiting part of the conversion is the calculation of the inverse
of $S$, which can be done with fast LAPACK routines.

\section{Conclusion}

\begin{figure}
\includegraphics[width=\columnwidth]{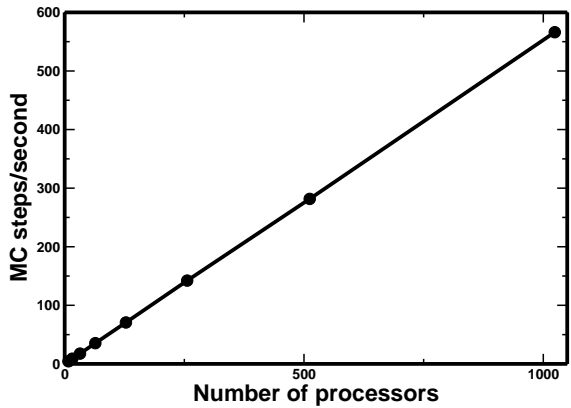}
\caption{Scaling of QWalk code over processors in Monte Carlo steps per second.  The system is a 
2x2x2 cell of BaTiO$_3$ with 320 electrons and one walker per node, on the San Diego Supercomputing 
Center DataStar machine.  This is VMC; DMC is very much the
same, because of the constant walker algorithm.  This is close to a worst-case scenario for QMC, 
since the run was only approximately 40 seconds long.}
\label{fig:qwalk_scaling}
\end{figure}

QWalk is a step forward in creating a state of the art, usable, and extensible program
for performing Quantum Monte Carlo calculations on electronic systems.  
It is able to handle medium to large systems of electrons; the 
maximum size is mostly limited by the available computer time.
It works in parallel very efficiently (Fig.~\ref{fig:qwalk_scaling}),
so it can take advantage of large clusters, multi-core computers, etc.
Since QWalk is available without charge and under the GNU Public license,
it is hoped that it will help bring both development and use of Quantum 
Monte Carlo methods to a wide audience. Due to its modular form it is 
straightforward to expand the QWalk's applicability to quantum systems
beyond the electron-ion Hamiltonians
in continuous space such as models of BEC/BCS condensates and other
quantum models. It is easy to modify the system module to incorporate 
other types of interactions and to expand the one-particle and pair
orbitals using the coded basis functions.   

We would like to extend our thanks to Zack Helms, David Sulock, Prasenjit Sen, 
Ji-Woo Lee, Jind\v rich Koloren\v c, Jeffrey Grossman, and Pavel Vagner for early testing of the code, 
and in the case of Zack Helms, Pavel Vagner, and David Sulock, contributions to some parts.  
This has been a long-term project and
funding has been provided by an NSF Graduate Research Fellowship for L. Wagner and
further by ONR-N00014-01-1-0408 grant, NSF grants DMR-0121361, DMR-0102668 and EAR-0530110. 

\begin{appendix}
\section{Adding a Module}
\label{app:qwalk_module}

We provide an example of how to add a new module.  In this case, we look at a Basis\_function
object, which has the fewest functions to fill in.  All modules can be added in exactly
the same way, differing only in what functions need to be defined.  Suppose we wish to 
add a Gaussian function $exp(-\alpha x^2)$. First we declare the
new module in a header file:
\begin{verbatim}
class Gaussian_basis:public Basis_function { 
public:
  //read the input
  void read(vector <string> & words); 

  //the distance at which the function is zero
  double cutoff(); 

  //The work functions.  Given a distance,
  //getVal returns the values
  //and getLap returns the values, first 
  //derivatives with respect to x,y, and z,
  //and the Laplacian
  void getVal(Array1 <doublevar> & r, 
              Array1 <doublevar> & vals);
  void getLap(Array1 <doublevar> & r, 
              Array2 <doublevar> & vals);
private:
  //put local variables here
  double alpha;
  double cut;
};
\end{verbatim}
Then we define the new functions:
\begin{verbatim}
Gaussian_basis::read(vector <string> & words) {
   unsigned int pos=0;
   if(!readvalue(words, pos,alpha, "ALPHA")) 
      error("Need ALPHA in gaussian basis");
   const double m=1e-18;
   cut=sqrt(-log(m/alpha));
}

Gaussian_basis::cutoff() { 
   return cut;
}

Gaussian_basis::getVal(Array1 <doublevar> & r,
                       Array1 <doublevar> & vals) { 
  //The basis function module can represent several 
  //functions, which are put into the vals array.
  //Here we only have one.
  //r is an array of form r,r^2,x,y,z
  vals(0)=exp(-alpha*r(1)); 
}

//getLap is omitted for space reasons
\end{verbatim}
The programmer then adds the source file to the Makefile and into a single if statement in the Basis\_function.cpp file.  The module can 
now be used anywhere another Basis\_funtion can be used.  All the modules follow this basic procedure, just with different
functions.

\end{appendix}

\bibliography{qwalk}

\begin{thebibliography}{49}
\expandafter\ifx\csname natexlab\endcsname\relax\def\natexlab#1{#1}\fi
\expandafter\ifx\csname bibnamefont\endcsname\relax
  \def\bibnamefont#1{#1}\fi
\expandafter\ifx\csname bibfnamefont\endcsname\relax
  \def\bibfnamefont#1{#1}\fi
\expandafter\ifx\csname citenamefont\endcsname\relax
  \def\citenamefont#1{#1}\fi
\expandafter\ifx\csname url\endcsname\relax
  \def\url#1{\texttt{#1}}\fi
\expandafter\ifx\csname urlprefix\endcsname\relax\def\urlprefix{URL }\fi
\providecommand{\bibinfo}[2]{#2}
\providecommand{\eprint}[2][]{\url{#2}}

\bibitem[{\citenamefont{Foulkes et~al.}(2001)\citenamefont{Foulkes, Mitas,
  Needs, and Rajagopal}}]{Foulkes_review}
\bibinfo{author}{\bibfnamefont{W.~M.~C.} \bibnamefont{Foulkes}},
  \bibinfo{author}{\bibfnamefont{L.}~\bibnamefont{Mitas}},
  \bibinfo{author}{\bibfnamefont{R.~J.} \bibnamefont{Needs}}, \bibnamefont{and}
  \bibinfo{author}{\bibfnamefont{G.}~\bibnamefont{Rajagopal}},
  \bibinfo{journal}{Rev Mod Phys} \textbf{\bibinfo{volume}{73}},
  \bibinfo{eid}{1} (\bibinfo{year}{2001}).

\bibitem[{\citenamefont{Grossman}(2002)}]{jeff_benchmark}
\bibinfo{author}{\bibfnamefont{J.}~\bibnamefont{Grossman}}, \bibinfo{journal}{J
  Chem. Phys.} \textbf{\bibinfo{volume}{117}}, \bibinfo{eid}{1434}
  (\bibinfo{year}{2002}).

\bibitem[{\citenamefont{Grossman and Mitas}(2005)}]{jeff_md}
\bibinfo{author}{\bibfnamefont{J.}~\bibnamefont{Grossman}} \bibnamefont{and}
  \bibinfo{author}{\bibfnamefont{L.}~\bibnamefont{Mitas}},
  \bibinfo{journal}{Phys. Rev. Lett.} \textbf{\bibinfo{volume}{94}},
  \bibinfo{eid}{056403} (\bibinfo{year}{2005}).

\bibitem[{\citenamefont{Bajdich
  et~al.}(2006{\natexlab{a}})\citenamefont{Bajdich, Mitas, Drobny, Wagner, and
  Schmidt}}]{michal_prl}
\bibinfo{author}{\bibfnamefont{M.}~\bibnamefont{Bajdich}},
  \bibinfo{author}{\bibfnamefont{L.}~\bibnamefont{Mitas}},
  \bibinfo{author}{\bibfnamefont{G.}~\bibnamefont{Drobny}},
  \bibinfo{author}{\bibfnamefont{L.}~\bibnamefont{Wagner}}, \bibnamefont{and}
  \bibinfo{author}{\bibfnamefont{K.}~\bibnamefont{Schmidt}},
  \bibinfo{journal}{Phys. Rev. Lett} \textbf{\bibinfo{volume}{96}},
  \bibinfo{eid}{130201} (\bibinfo{year}{2006}{\natexlab{a}}).

\bibitem[{\citenamefont{Bajdich et~al.}(2005)\citenamefont{Bajdich, Mitas,
  Drobny, and Wagner}}]{michal_prb}
\bibinfo{author}{\bibfnamefont{M.}~\bibnamefont{Bajdich}},
  \bibinfo{author}{\bibfnamefont{L.}~\bibnamefont{Mitas}},
  \bibinfo{author}{\bibfnamefont{G.}~\bibnamefont{Drobny}}, \bibnamefont{and}
  \bibinfo{author}{\bibfnamefont{L.}~\bibnamefont{Wagner}},
  \bibinfo{journal}{Phys. Rev. B} \textbf{\bibinfo{volume}{72}},
  \bibinfo{eid}{075131} (\bibinfo{year}{2005}).

\bibitem[{\citenamefont{Vagner et~al.}(2006)\citenamefont{Vagner, Mosko,
  Nemeth, Wagner, and Mitas}}]{pavel_1d}
\bibinfo{author}{\bibfnamefont{P.}~\bibnamefont{Vagner}},
  \bibinfo{author}{\bibfnamefont{M.}~\bibnamefont{Mosko}},
  \bibinfo{author}{\bibfnamefont{R.}~\bibnamefont{Nemeth}},
  \bibinfo{author}{\bibfnamefont{L.}~\bibnamefont{Wagner}}, \bibnamefont{and}
  \bibinfo{author}{\bibfnamefont{L.}~\bibnamefont{Mitas}},
  \bibinfo{journal}{Physica E} p. \bibinfo{pages}{350} (\bibinfo{year}{2006}).

\bibitem[{\citenamefont{Wagner and Mitas}(2007)}]{lucas_tmo_mol}
\bibinfo{author}{\bibfnamefont{L.~K.} \bibnamefont{Wagner}} \bibnamefont{and}
  \bibinfo{author}{\bibfnamefont{L.}~\bibnamefont{Mitas}},
  \bibinfo{journal}{The Journal of Chemical Physics}
  \textbf{\bibinfo{volume}{126}}, \bibinfo{eid}{034105}
  (pages~\bibinfo{numpages}{5}) (\bibinfo{year}{2007}),
  \urlprefix\url{http://link.aip.org/link/?JCP/126/034105/1}.

\bibitem[{\citenamefont{Bajdich
  et~al.}(2006{\natexlab{b}})\citenamefont{Bajdich, Mitas, Wagner, and
  Schmidt}}]{michal_pfaff_prb}
\bibinfo{author}{\bibfnamefont{M.}~\bibnamefont{Bajdich}},
  \bibinfo{author}{\bibfnamefont{L.}~\bibnamefont{Mitas}},
  \bibinfo{author}{\bibfnamefont{L.}~\bibnamefont{Wagner}}, \bibnamefont{and}
  \bibinfo{author}{\bibfnamefont{K.}~\bibnamefont{Schmidt}},
  \bibinfo{journal}{submitted to Phys. Rev. B} pp.
  \bibinfo{pages}{arxiv:cond--mat/0610850}
  (\bibinfo{year}{2006}{\natexlab{b}}).

\bibitem[{\citenamefont{Metropolis et~al.}(1953)\citenamefont{Metropolis,
  Rosenbluth, Rosenbluth, Teller, and Teller}}]{metropolis}
\bibinfo{author}{\bibfnamefont{N.}~\bibnamefont{Metropolis}},
  \bibinfo{author}{\bibfnamefont{A.}~\bibnamefont{Rosenbluth}},
  \bibinfo{author}{\bibfnamefont{M.}~\bibnamefont{Rosenbluth}},
  \bibinfo{author}{\bibfnamefont{A.}~\bibnamefont{Teller}}, \bibnamefont{and}
  \bibinfo{author}{\bibfnamefont{E.}~\bibnamefont{Teller}},
  \bibinfo{journal}{J. Chem. Phys.} \textbf{\bibinfo{volume}{21}},
  \bibinfo{pages}{1087} (\bibinfo{year}{1953}).

\bibitem[{\citenamefont{Hastings}(1970)}]{hastings}
\bibinfo{author}{\bibfnamefont{W.}~\bibnamefont{Hastings}},
  \bibinfo{journal}{Biometrika} \textbf{\bibinfo{volume}{57}},
  \bibinfo{pages}{97} (\bibinfo{year}{1970}).

\bibitem[{\citenamefont{Bressanini et~al.}(2004)\citenamefont{Bressanini,
  Morosi, Tarasco, and Mira}}]{bressanini:3446}
\bibinfo{author}{\bibfnamefont{D.}~\bibnamefont{Bressanini}},
  \bibinfo{author}{\bibfnamefont{G.}~\bibnamefont{Morosi}},
  \bibinfo{author}{\bibfnamefont{S.}~\bibnamefont{Tarasco}}, \bibnamefont{and}
  \bibinfo{author}{\bibfnamefont{A.}~\bibnamefont{Mira}}, \bibinfo{journal}{The
  Journal of Chemical Physics} \textbf{\bibinfo{volume}{121}},
  \bibinfo{pages}{3446} (\bibinfo{year}{2004}),
  \urlprefix\url{http://link.aip.org/link/?JCP/121/3446/1}.

\bibitem[{\citenamefont{Umrigar et~al.}(1993)\citenamefont{Umrigar,
  Nightingale, and Runge}}]{unr}
\bibinfo{author}{\bibfnamefont{C.}~\bibnamefont{Umrigar}},
  \bibinfo{author}{\bibfnamefont{M.}~\bibnamefont{Nightingale}},
  \bibnamefont{and} \bibinfo{author}{\bibfnamefont{K.}~\bibnamefont{Runge}},
  \bibinfo{journal}{J. Chem. Phys.} \textbf{\bibinfo{volume}{99}},
  \bibinfo{eid}{2865} (\bibinfo{year}{1993}).

\bibitem[{\citenamefont{Anderson}(1975)}]{jbanderson75}
\bibinfo{author}{\bibfnamefont{J.~B.} \bibnamefont{Anderson}},
  \bibinfo{journal}{The Journal of Chemical Physics}
  \textbf{\bibinfo{volume}{63}}, \bibinfo{pages}{1499} (\bibinfo{year}{1975}),
  \urlprefix\url{http://link.aip.org/link/?JCP/63/1499/1}.

\bibitem[{\citenamefont{Anderson}(1976)}]{jbanderson76}
\bibinfo{author}{\bibfnamefont{J.~B.} \bibnamefont{Anderson}},
  \bibinfo{journal}{The Journal of Chemical Physics}
  \textbf{\bibinfo{volume}{65}}, \bibinfo{pages}{4121} (\bibinfo{year}{1976}),
  \urlprefix\url{http://link.aip.org/link/?JCP/65/4121/1}.

\bibitem[{\citenamefont{Moskowitz et~al.}(1982)\citenamefont{Moskowitz,
  Schmidt, Lee, and Kalos}}]{moskowitz82}
\bibinfo{author}{\bibfnamefont{J.~W.} \bibnamefont{Moskowitz}},
  \bibinfo{author}{\bibfnamefont{K.~E.} \bibnamefont{Schmidt}},
  \bibinfo{author}{\bibfnamefont{M.~A.} \bibnamefont{Lee}}, \bibnamefont{and}
  \bibinfo{author}{\bibfnamefont{M.~H.} \bibnamefont{Kalos}},
  \bibinfo{journal}{The Journal of Chemical Physics}
  \textbf{\bibinfo{volume}{77}}, \bibinfo{pages}{349} (\bibinfo{year}{1982}),
  \urlprefix\url{http://link.aip.org/link/?JCP/77/349/1}.

\bibitem[{\citenamefont{Reynolds
  et~al.}(1982{\natexlab{a}})\citenamefont{Reynolds, Ceperley, Alder, and
  \mbox{Lester Jr.}}}]{reynolds82}
\bibinfo{author}{\bibfnamefont{P.~J.} \bibnamefont{Reynolds}},
  \bibinfo{author}{\bibfnamefont{D.~M.} \bibnamefont{Ceperley}},
  \bibinfo{author}{\bibfnamefont{B.~J.} \bibnamefont{Alder}}, \bibnamefont{and}
  \bibinfo{author}{\bibfnamefont{W.~A.} \bibnamefont{\mbox{Lester Jr.}}},
  \bibinfo{journal}{The Journal of Chemical Physics}
  \textbf{\bibinfo{volume}{77}}, \bibinfo{pages}{5593}
  (\bibinfo{year}{1982}{\natexlab{a}}),
  \urlprefix\url{http://link.aip.org/link/?JCP/77/5593/1}.

\bibitem[{\citenamefont{Ceperley and Adler}(1980)}]{ceperley_adler}
\bibinfo{author}{\bibfnamefont{D.}~\bibnamefont{Ceperley}} \bibnamefont{and}
  \bibinfo{author}{\bibfnamefont{B.}~\bibnamefont{Adler}},
  \bibinfo{journal}{Phys. Rev. Lett.} \textbf{\bibinfo{volume}{45}},
  \bibinfo{eid}{566} (\bibinfo{year}{1980}).

\bibitem[{\citenamefont{Baroni and Moroni}(1999)}]{Baroni_RMC}
\bibinfo{author}{\bibfnamefont{S.}~\bibnamefont{Baroni}} \bibnamefont{and}
  \bibinfo{author}{\bibfnamefont{S.}~\bibnamefont{Moroni}},
  \bibinfo{journal}{Phys. Rev. Lett.} \textbf{\bibinfo{volume}{82}},
  \bibinfo{eid}{4745} (\bibinfo{year}{1999}).

\bibitem[{\citenamefont{Pierleoni and Ceperley}(2005)}]{pierleoni_rmc}
\bibinfo{author}{\bibfnamefont{C.}~\bibnamefont{Pierleoni}} \bibnamefont{and}
  \bibinfo{author}{\bibfnamefont{D.~M.} \bibnamefont{Ceperley}},
  \bibinfo{journal}{ChemPhysChem} \textbf{\bibinfo{volume}{6}},
  \bibinfo{eid}{1872} (\bibinfo{year}{2005}).

\bibitem[{\citenamefont{Umrigar et~al.}(1988)\citenamefont{Umrigar, Wilson, and
  Wilkins}}]{umrigar_varopt}
\bibinfo{author}{\bibfnamefont{C.}~\bibnamefont{Umrigar}},
  \bibinfo{author}{\bibfnamefont{K.}~\bibnamefont{Wilson}}, \bibnamefont{and}
  \bibinfo{author}{\bibfnamefont{J.}~\bibnamefont{Wilkins}},
  \bibinfo{journal}{Phys. Rev. Lett.} \textbf{\bibinfo{volume}{60}},
  \bibinfo{eid}{1719} (\bibinfo{year}{1988}).

\bibitem[{\citenamefont{Umrigar and Filippi}(2005)}]{umrigar_optimization2}
\bibinfo{author}{\bibfnamefont{C.~J.} \bibnamefont{Umrigar}} \bibnamefont{and}
  \bibinfo{author}{\bibfnamefont{C.}~\bibnamefont{Filippi}},
  \bibinfo{journal}{Phys Rev Lett} \textbf{\bibinfo{volume}{94}},
  \bibinfo{eid}{150201} (\bibinfo{year}{2005}).

\bibitem[{\citenamefont{Filippi and Umrigar}(2000)}]{filippi_force}
\bibinfo{author}{\bibfnamefont{C.}~\bibnamefont{Filippi}} \bibnamefont{and}
  \bibinfo{author}{\bibfnamefont{C.~J.} \bibnamefont{Umrigar}},
  \bibinfo{journal}{Phys Rev B} \textbf{\bibinfo{volume}{61}},
  \bibinfo{eid}{R16291} (\bibinfo{year}{2000}).

\bibitem[{\citenamefont{Chiesa et~al.}(2006)\citenamefont{Chiesa, Ceperley,
  Martin, and Holzmann}}]{chiesa_finite_size}
\bibinfo{author}{\bibfnamefont{S.}~\bibnamefont{Chiesa}},
  \bibinfo{author}{\bibfnamefont{D.~M.} \bibnamefont{Ceperley}},
  \bibinfo{author}{\bibfnamefont{R.~M.} \bibnamefont{Martin}},
  \bibnamefont{and} \bibinfo{author}{\bibfnamefont{M.}~\bibnamefont{Holzmann}},
  \bibinfo{journal}{Physical Review Letters} \textbf{\bibinfo{volume}{97}},
  \bibinfo{eid}{076404} (pages~\bibinfo{numpages}{4}) (\bibinfo{year}{2006}),
  \urlprefix\url{http://link.aps.org/abstract/PRL/v97/e076404}.

\bibitem[{\citenamefont{Williamson et~al.}(1997)\citenamefont{Williamson,
  Rajagopal, Needs, Fraser, Foulkes, Wang, and Chou}}]{cambridge_mpc}
\bibinfo{author}{\bibfnamefont{A.~J.} \bibnamefont{Williamson}},
  \bibinfo{author}{\bibfnamefont{G.}~\bibnamefont{Rajagopal}},
  \bibinfo{author}{\bibfnamefont{R.~J.} \bibnamefont{Needs}},
  \bibinfo{author}{\bibfnamefont{L.~M.} \bibnamefont{Fraser}},
  \bibinfo{author}{\bibfnamefont{W.~M.~C.} \bibnamefont{Foulkes}},
  \bibinfo{author}{\bibfnamefont{Y.}~\bibnamefont{Wang}}, \bibnamefont{and}
  \bibinfo{author}{\bibfnamefont{M.-Y.} \bibnamefont{Chou}},
  \bibinfo{journal}{Physical Review B (Condensed Matter)}
  \textbf{\bibinfo{volume}{55}}, \bibinfo{pages}{R4851} (\bibinfo{year}{1997}),
  \urlprefix\url{http://link.aps.org/abstract/PRB/v55/pR4851}.

\bibitem[{\citenamefont{Barnett et~al.}(2001)\citenamefont{Barnett, Sun, and
  Lester}}]{Lester_psp}
\bibinfo{author}{\bibfnamefont{R.}~\bibnamefont{Barnett}},
  \bibinfo{author}{\bibfnamefont{Z.}~\bibnamefont{Sun}}, \bibnamefont{and}
  \bibinfo{author}{\bibfnamefont{W.}~\bibnamefont{Lester}},
  \bibinfo{journal}{J. Chem. Phys.} \textbf{\bibinfo{volume}{114}},
  \bibinfo{pages}{7790} (\bibinfo{year}{2001}).

\bibitem[{\citenamefont{Lee et~al.}(2000)\citenamefont{Lee, Kent, Towler,
  Needs, and Rajagopal}}]{Lee_psp}
\bibinfo{author}{\bibfnamefont{Y.}~\bibnamefont{Lee}},
  \bibinfo{author}{\bibfnamefont{P.}~\bibnamefont{Kent}},
  \bibinfo{author}{\bibfnamefont{M.}~\bibnamefont{Towler}},
  \bibinfo{author}{\bibfnamefont{R.}~\bibnamefont{Needs}}, \bibnamefont{and}
  \bibinfo{author}{\bibfnamefont{G.}~\bibnamefont{Rajagopal}},
  \bibinfo{journal}{Phys Rev B and private communication}
  \textbf{\bibinfo{volume}{62}}, \bibinfo{eid}{13347} (\bibinfo{year}{2000}).

\bibitem[{\citenamefont{Trail and Needs}(2005)}]{Trail_psp}
\bibinfo{author}{\bibfnamefont{J.}~\bibnamefont{Trail}} \bibnamefont{and}
  \bibinfo{author}{\bibfnamefont{R.}~\bibnamefont{Needs}}, \bibinfo{journal}{J.
  Chem. Phys.} \textbf{\bibinfo{volume}{122}}, \bibinfo{pages}{174109}
  (\bibinfo{year}{2005}).

\bibitem[{\citenamefont{Burkatzki et~al.}(2007)\citenamefont{Burkatzki,
  Filippi, and Dolg}}]{Dolg_psp_qmc}
\bibinfo{author}{\bibfnamefont{M.}~\bibnamefont{Burkatzki}},
  \bibinfo{author}{\bibfnamefont{C.}~\bibnamefont{Filippi}}, \bibnamefont{and}
  \bibinfo{author}{\bibfnamefont{M.}~\bibnamefont{Dolg}}, \bibinfo{journal}{The
  Journal of Chemical Physics} \textbf{\bibinfo{volume}{126}},
  \bibinfo{eid}{234105} (pages~\bibinfo{numpages}{8}) (\bibinfo{year}{2007}),
  \urlprefix\url{http://link.aip.org/link/?JCP/126/234105/1}.

\bibitem[{\citenamefont{Casula et~al.}(2004)\citenamefont{Casula, Attaccalite,
  and Sorella}}]{casula-2004-121}
\bibinfo{author}{\bibfnamefont{M.}~\bibnamefont{Casula}},
  \bibinfo{author}{\bibfnamefont{C.}~\bibnamefont{Attaccalite}},
  \bibnamefont{and} \bibinfo{author}{\bibfnamefont{S.}~\bibnamefont{Sorella}},
  \bibinfo{journal}{J.CHEM.PHYS.} \textbf{\bibinfo{volume}{121}},
  \bibinfo{pages}{7110} (\bibinfo{year}{2004}),
  \urlprefix\url{http://www.citebase.org/cgi-bin/citations?id=oai:arXiv.org:co%
nd-mat/0409644}.

\bibitem[{\citenamefont{Bajdich}(2007)}]{mb_thesis}
\bibinfo{author}{\bibfnamefont{M.}~\bibnamefont{Bajdich}}, Ph.D. thesis,
  \bibinfo{school}{North Carolina State University} (\bibinfo{year}{2007}),
  \urlprefix\url{http://altair.physics.ncsu.edu/bajdich/Phd_thesis_m_bajdich.p%
df}.

\bibitem[{\citenamefont{Feynman and Cohen}(1956)}]{feynman}
\bibinfo{author}{\bibfnamefont{R.~P.} \bibnamefont{Feynman}} \bibnamefont{and}
  \bibinfo{author}{\bibfnamefont{M.}~\bibnamefont{Cohen}},
  \bibinfo{journal}{Phys. Rev.} \textbf{\bibinfo{volume}{102}},
  \bibinfo{pages}{1189} (\bibinfo{year}{1956}).

\bibitem[{\citenamefont{Schmidt et~al.}(1981)\citenamefont{Schmidt, Lee, Kalos,
  and Chester}}]{schmidt_bf}
\bibinfo{author}{\bibfnamefont{K.~E.} \bibnamefont{Schmidt}},
  \bibinfo{author}{\bibfnamefont{M.~A.} \bibnamefont{Lee}},
  \bibinfo{author}{\bibfnamefont{M.~H.} \bibnamefont{Kalos}}, \bibnamefont{and}
  \bibinfo{author}{\bibfnamefont{G.~V.} \bibnamefont{Chester}},
  \bibinfo{journal}{Phys. Rev. Lett.} \textbf{\bibinfo{volume}{47}},
  \bibinfo{pages}{807} (\bibinfo{year}{1981}).

\bibitem[{\citenamefont{Panoff and Carlson}(1989)}]{panoff}
\bibinfo{author}{\bibfnamefont{R.~M.} \bibnamefont{Panoff}} \bibnamefont{and}
  \bibinfo{author}{\bibfnamefont{J.}~\bibnamefont{Carlson}},
  \bibinfo{journal}{Phys. Rev. Lett.} \textbf{\bibinfo{volume}{62}},
  \bibinfo{pages}{1130} (\bibinfo{year}{1989}).

\bibitem[{\citenamefont{Moskowitz and Schmidt}(1992)}]{moskowitz}
\bibinfo{author}{\bibfnamefont{J.~W.} \bibnamefont{Moskowitz}}
  \bibnamefont{and} \bibinfo{author}{\bibfnamefont{K.~E.}
  \bibnamefont{Schmidt}}, \bibinfo{journal}{The Journal of Chemical Physics}
  \textbf{\bibinfo{volume}{97}}, \bibinfo{pages}{3382} (\bibinfo{year}{1992}),
  \urlprefix\url{http://link.aip.org/link/?JCP/97/3382/1}.

\bibitem[{\citenamefont{Kwon et~al.}(1993)\citenamefont{Kwon, Ceperley, and
  Martin}}]{kwon1}
\bibinfo{author}{\bibfnamefont{Y.}~\bibnamefont{Kwon}},
  \bibinfo{author}{\bibfnamefont{D.~M.} \bibnamefont{Ceperley}},
  \bibnamefont{and} \bibinfo{author}{\bibfnamefont{R.~M.}
  \bibnamefont{Martin}}, \bibinfo{journal}{Phys. Rev. B}
  \textbf{\bibinfo{volume}{48}}, \bibinfo{pages}{12037} (\bibinfo{year}{1993}).

\bibitem[{\citenamefont{Kwon et~al.}(1994)\citenamefont{Kwon, Ceperley, and
  Martin}}]{kwon2}
\bibinfo{author}{\bibfnamefont{Y.}~\bibnamefont{Kwon}},
  \bibinfo{author}{\bibfnamefont{D.~M.} \bibnamefont{Ceperley}},
  \bibnamefont{and} \bibinfo{author}{\bibfnamefont{R.~M.}
  \bibnamefont{Martin}}, \bibinfo{journal}{Phys. Rev. B}
  \textbf{\bibinfo{volume}{50}}, \bibinfo{pages}{1684} (\bibinfo{year}{1994}).

\bibitem[{\citenamefont{Kwon et~al.}(1996)\citenamefont{Kwon, Ceperley, and
  Martin}}]{kwon3}
\bibinfo{author}{\bibfnamefont{Y.}~\bibnamefont{Kwon}},
  \bibinfo{author}{\bibfnamefont{D.~M.} \bibnamefont{Ceperley}},
  \bibnamefont{and} \bibinfo{author}{\bibfnamefont{R.~M.}
  \bibnamefont{Martin}}, \bibinfo{journal}{Phys. Rev. B}
  \textbf{\bibinfo{volume}{53}}, \bibinfo{pages}{7376} (\bibinfo{year}{1996}).

\bibitem[{\citenamefont{Drummond et~al.}(2006)\citenamefont{Drummond, Rios, Ma,
  Trail, Spink, Towler, and Needs}}]{drummond_bf}
\bibinfo{author}{\bibfnamefont{N.~D.} \bibnamefont{Drummond}},
  \bibinfo{author}{\bibfnamefont{P.~L.} \bibnamefont{Rios}},
  \bibinfo{author}{\bibfnamefont{A.}~\bibnamefont{Ma}},
  \bibinfo{author}{\bibfnamefont{J.~R.} \bibnamefont{Trail}},
  \bibinfo{author}{\bibfnamefont{G.~G.} \bibnamefont{Spink}},
  \bibinfo{author}{\bibfnamefont{M.~D.} \bibnamefont{Towler}},
  \bibnamefont{and} \bibinfo{author}{\bibfnamefont{R.~J.} \bibnamefont{Needs}},
  \bibinfo{journal}{The Journal of Chemical Physics}
  \textbf{\bibinfo{volume}{124}}, \bibinfo{eid}{224104}
  (pages~\bibinfo{numpages}{6}) (\bibinfo{year}{2006}),
  \urlprefix\url{http://link.aip.org/link/?JCP/124/224104/1}.

\bibitem[{\citenamefont{Rios et~al.}(2006)\citenamefont{Rios, Ma, Drummond,
  Towler, and Needs}}]{rios_bf}
\bibinfo{author}{\bibfnamefont{P.~L.} \bibnamefont{Rios}},
  \bibinfo{author}{\bibfnamefont{A.}~\bibnamefont{Ma}},
  \bibinfo{author}{\bibfnamefont{N.~D.} \bibnamefont{Drummond}},
  \bibinfo{author}{\bibfnamefont{M.~D.} \bibnamefont{Towler}},
  \bibnamefont{and} \bibinfo{author}{\bibfnamefont{R.~J.} \bibnamefont{Needs}},
  \bibinfo{journal}{Physical Review E (Statistical, Nonlinear, and Soft Matter
  Physics)} \textbf{\bibinfo{volume}{74}}, \bibinfo{eid}{066701}
  (pages~\bibinfo{numpages}{15}) (\bibinfo{year}{2006}),
  \urlprefix\url{http://link.aps.org/abstract/PRE/v74/e066701}.

\bibitem[{\citenamefont{Schmidt et~al.}(1993)\citenamefont{Schmidt, Baldridge,
  Boatz, Elbert, Gordon, Jensen, Koseki, Matsunaga, Nguyen, Su
  et~al.}}]{gamess}
\bibinfo{author}{\bibfnamefont{M.~W.} \bibnamefont{Schmidt}},
  \bibinfo{author}{\bibfnamefont{K.~K.} \bibnamefont{Baldridge}},
  \bibinfo{author}{\bibfnamefont{J.~A.} \bibnamefont{Boatz}},
  \bibinfo{author}{\bibfnamefont{S.~T.} \bibnamefont{Elbert}},
  \bibinfo{author}{\bibfnamefont{M.~S.} \bibnamefont{Gordon}},
  \bibinfo{author}{\bibfnamefont{J.~H.} \bibnamefont{Jensen}},
  \bibinfo{author}{\bibfnamefont{S.}~\bibnamefont{Koseki}},
  \bibinfo{author}{\bibfnamefont{N.}~\bibnamefont{Matsunaga}},
  \bibinfo{author}{\bibfnamefont{K.~A.} \bibnamefont{Nguyen}},
  \bibinfo{author}{\bibfnamefont{S.~J.} \bibnamefont{Su}},
  \bibnamefont{et~al.}, \bibinfo{journal}{J. Comput Chem}
  \textbf{\bibinfo{volume}{14}}, \bibinfo{eid}{1347} (\bibinfo{year}{1993}).

\bibitem[{\citenamefont{Saunders et~al.}(2003)\citenamefont{Saunders, Dovesi,
  Roetti, Orlando, Zicovich-Wilson, Harrison, Doll, Civalleri, Bush, D’Arco
  et~al.}}]{crystal}
\bibinfo{author}{\bibfnamefont{V.}~\bibnamefont{Saunders}},
  \bibinfo{author}{\bibfnamefont{R.}~\bibnamefont{Dovesi}},
  \bibinfo{author}{\bibfnamefont{C.}~\bibnamefont{Roetti}},
  \bibinfo{author}{\bibfnamefont{R.}~\bibnamefont{Orlando}},
  \bibinfo{author}{\bibfnamefont{C.}~\bibnamefont{Zicovich-Wilson}},
  \bibinfo{author}{\bibfnamefont{N.}~\bibnamefont{Harrison}},
  \bibinfo{author}{\bibfnamefont{K.}~\bibnamefont{Doll}},
  \bibinfo{author}{\bibfnamefont{B.}~\bibnamefont{Civalleri}},
  \bibinfo{author}{\bibfnamefont{I.}~\bibnamefont{Bush}},
  \bibinfo{author}{\bibfnamefont{P.}~\bibnamefont{D’Arco}},
  \bibnamefont{et~al.} (\bibinfo{year}{2003}).

\bibitem[{\citenamefont{Jos\'{e} M~Soler and
  S\'{a}nchez-Portal}(2002)}]{siesta}
\bibinfo{author}{\bibfnamefont{J.~D. G. A. G. J. J. P.~O.}
  \bibnamefont{Jos\'{e} M~Soler}, \bibfnamefont{Emilio~Artacho}}
  \bibnamefont{and}
  \bibinfo{author}{\bibfnamefont{D.}~\bibnamefont{S\'{a}nchez-Portal}},
  \bibinfo{journal}{Journal of Physics: Condensed Matter}
  \textbf{\bibinfo{volume}{14}}, \bibinfo{pages}{2745} (\bibinfo{year}{2002}),
  \urlprefix\url{http://stacks.iop.org/0953-8984/14/2745}.

\bibitem[{\citenamefont{Gygi}(2005)}]{GP}
\bibinfo{author}{\bibfnamefont{F.}~\bibnamefont{Gygi}} (\bibinfo{year}{2005}).

\bibitem[{\citenamefont{Mitas et~al.}(1991)\citenamefont{Mitas, Shirley, and
  Ceperley}}]{lubos_psp}
\bibinfo{author}{\bibfnamefont{L.}~\bibnamefont{Mitas}},
  \bibinfo{author}{\bibfnamefont{E.}~\bibnamefont{Shirley}}, \bibnamefont{and}
  \bibinfo{author}{\bibfnamefont{D.}~\bibnamefont{Ceperley}},
  \bibinfo{journal}{J. Chem Phys} \textbf{\bibinfo{volume}{95}},
  \bibinfo{eid}{3467} (\bibinfo{year}{1991}).

\bibitem[{\citenamefont{Reynolds
  et~al.}(1982{\natexlab{b}})\citenamefont{Reynolds, Ceperley, Alder, and
  {Lester, Jr.}}}]{reynolds}
\bibinfo{author}{\bibfnamefont{P.}~\bibnamefont{Reynolds}},
  \bibinfo{author}{\bibfnamefont{D.}~\bibnamefont{Ceperley}},
  \bibinfo{author}{\bibfnamefont{B.}~\bibnamefont{Alder}}, \bibnamefont{and}
  \bibinfo{author}{\bibfnamefont{W.}~\bibnamefont{{Lester, Jr.}}},
  \bibinfo{journal}{J. Chem. Phys.} \textbf{\bibinfo{volume}{77}},
  \bibinfo{eid}{5593} (\bibinfo{year}{1982}{\natexlab{b}}).

\bibitem[{\citenamefont{Honeycutt}(1992)}]{srk}
\bibinfo{author}{\bibfnamefont{R.~L.} \bibnamefont{Honeycutt}},
  \bibinfo{journal}{Phys. Rev. A} \textbf{\bibinfo{volume}{45}},
  \bibinfo{pages}{600} (\bibinfo{year}{1992}).

\bibitem[{\citenamefont{Manten and L\"uchow}(2003)}]{luechow_linear}
\bibinfo{author}{\bibfnamefont{S.}~\bibnamefont{Manten}} \bibnamefont{and}
  \bibinfo{author}{\bibfnamefont{A.}~\bibnamefont{L\"uchow}},
  \bibinfo{journal}{J. Chem. Phys.} \textbf{\bibinfo{volume}{119}},
  \bibinfo{eid}{1307} (\bibinfo{year}{2003}).

\bibitem[{\citenamefont{Kwon et~al.}(1998)\citenamefont{Kwon, Ceperley, and
  Martin}}]{kwon4}
\bibinfo{author}{\bibfnamefont{Y.}~\bibnamefont{Kwon}},
  \bibinfo{author}{\bibfnamefont{D.~M.} \bibnamefont{Ceperley}},
  \bibnamefont{and} \bibinfo{author}{\bibfnamefont{R.~M.}
  \bibnamefont{Martin}}, \bibinfo{journal}{Phys. Rev. B}
  \textbf{\bibinfo{volume}{58}}, \bibinfo{pages}{6800} (\bibinfo{year}{1998}).

\bibitem[{\citenamefont{Holzmann et~al.}(2003)\citenamefont{Holzmann, Ceperley,
  Pierleoni, and Esler}}]{markus}
\bibinfo{author}{\bibfnamefont{M.}~\bibnamefont{Holzmann}},
  \bibinfo{author}{\bibfnamefont{D.~M.} \bibnamefont{Ceperley}},
  \bibinfo{author}{\bibfnamefont{C.}~\bibnamefont{Pierleoni}},
  \bibnamefont{and} \bibinfo{author}{\bibfnamefont{K.}~\bibnamefont{Esler}},
  \bibinfo{journal}{Physical Review E (Statistical, Nonlinear, and Soft Matter
  Physics)} \textbf{\bibinfo{volume}{68}}, \bibinfo{eid}{046707}
  (pages~\bibinfo{numpages}{15}) (\bibinfo{year}{2003}),
  \urlprefix\url{http://link.aps.org/abstract/PRE/v68/e046707}.

\end{thebibliography}
\nocite{*}

\end{document}